\documentclass[superscript address,twocolumn]{revtex4-2}

\usepackage{graphics}
\usepackage{graphicx}
\usepackage{bm}
\usepackage{amsmath}
\usepackage{mathtools}
\usepackage{color}

\begin{document}	
	\title{A simple quantum picture of the relativistic Doppler effect}
	
	\author{Daniel Hodgson}
	\affiliation{School of Physics and Astronomy, University of Leeds, Leeds, UK, LS2 9JT}
	\author{Sara Kanzi}
	\affiliation{Faculty of Engineering, Final International University, North Cyprus Via Mersin 10, Kyrenia, 99370, Turkey}
	\author{Almut Beige}
	\affiliation{School of Physics and Astronomy, University of Leeds, Leeds, UK, LS2 9JT}
	
	\begin{abstract}
		The relativistic Doppler effect comes from the fact that observers in different inertial reference frames experience space and time differently, while the speed of light remains always the same. Consequently, a wave packet of light exhibits different frequencies, wavelengths, and amplitudes. In this paper, we present a local approach to the relativistic Doppler effect based on relativity, spatial and time translational symmetries, and energy conservation. Afterwards we investigate the implications of the relativistic Doppler effect for the quantum state transformations of wave packets of light and show that a local photon is a local photon at the same point in the spacetime diagram in all inertial frames.
	\end{abstract}
	
	\maketitle
	
	\section{Introduction}
	
	When a moving car beeps its horn, the driver and a bystander on the pavement hear the sound at different frequencies and observe different wavelengths. This change, resulting from the relative motion of the driver and the bystander, is known as the Doppler effect \cite{classicalDoppler,classicalDoppler2} and is well understood in classical physics. For example, the frequency heard by the resting observer depends on the speed of the car relative to the pavement and the original frequency of the signal. Due to its simplicity, the Doppler effect has already found a wide range of applications, including the policing of speed limit violations by irresponsible drivers. The relativistic Doppler effect \cite{Jones1939,Otting1939,relDoppler,Kaivola1985, Mandelberg1962, Olin1973,Schachinger} also accounts for differences in how observers experience space and time. Observers in different inertial reference frames which move at a relative speed close to the speed of light receive signals which differ not only in frequency and wavelength but also in amplitude.
	
	According to Einstein's principle of relativity \cite{Einstein,timedilation,Crouse:2018slq,GWINNER2005,Saathoff2003,Hafele,Unnikrishnan:2004nj}, there is no privileged frame of reference. The same physical laws apply in all reference frames if these move with respect to each other at constant velocity. For example, wave packets of light with a well-defined direction of propagation move at the speed of light $c$ in any reference frame. For completeness, let us point out that some authors debate whether this assumption is true or not \cite{doubters}. For example, it is believed that some effects, such as the experimentally verified Sagnac effect \cite{Sagnac1, Sagnac2}, are best understood in terms of anisotropies of the speed of light \cite{Sagnac3}. Moreover, some experiments that have been designed to disprove the existence of an aether may have been misinterpreted \cite{Khan1,Khan2}. Other experiments again claim to verify the constancy of the speed of light with high accuracy \cite{experiments,experiments2}. Here we notice that any physical theory that involves space and time requires a way of measuring both using clocks and meters and simply assume in the following that all clocks and meters are calibrated such that light travels at the same speed in all directions in all reference frames. 
	
	Recently, our group discussed and promoted the possible quantisation of the electromagnetic (EM) field in position space \cite{Ali,Southall:2019zdm,Hodgson:2021eyb,Hodgson2}. Starting from the assumption that the basic building blocks of light are localised photons\textemdash so-called {\it blips} (bosons localised in position)\textemdash with well-defined positions, polarisations and directions of propagation, we derived a Hamiltonian that generates their dynamics as well as electric and magnetic field observables for the calculation of expectation values. As a first application of our local photon approach, we constructed locally-acting mirror Hamiltonians for describing light scattering by partially transparent interfaces \cite{Southall:2019zdm,AMC}. However, our approach differs from previous field quantisation schemes (cf.~e.g.~Ref.~\cite{Bennett} and references therein) and requires a doubling of the Hilbert space of the EM field by the inclusion of positive as well as negative frequency photons. The main purpose of this manuscript is to verify the consistency of our generalisation of standard quantum optics approaches with the well-known Doppler effect \cite{relDoppler,Jones1939, Kaivola1985, Mandelberg1962, Olin1973}.
	
	In addition to demonstrating this consistency, we show that a local approach has many advantages and offers new insight. For example, as we shall see below, it can easily accommodate spatial and time translational symmetries in a straightforward way. In the following, we derive the relativistic Doppler effect with only a minimum of assumptions, and, as we shall see below, all results presented here are consistent with the existing literature \cite{Fang2021,Ran2016,Li2016,Klacka1992,Giuliani2014,Navia:2006hk,Jiang2018,GUO2022,DASANNACHARYA1944}. The main new result of our investigation will be the derivation of a relationship between local photons in different inertial reference frames. More concretely, it will be shown that a local photon is seen as a local photon by all observers at the same point in the spacetime diagram. We are therefore confident that our local approach will pave the way for systematic studies of even more complex scenarios, like the Unruh effect \cite{Unruh1976,Unruh2} and quantum electrodynamics in reference frames with time-varying accelerations without the need for approximations, such as the usual assumption of a flat spacetime \cite{Maybee}. Moreover, some of the insights obtained here might have applications in relativistic quantum information \cite{Bruschi2,Bruschi4, Perdigues2008,Bruschi3, Ralph:2011hp,Friis:2012cx,Ursin2009, Alsing:2012wf,Rideout:2012jb}. 
	
	In the following we review the basic assumptions made in the derivation of the relativistic Doppler effect. Suppose an observer\textemdash let us call her Alice (A)\textemdash is watching a wave packet of light with a well-defined direction of propagation $s$ and a well-defined polarisation $\lambda$ travel along the $x$ axis. Since this wave packet travels at the speed of light, its electric field amplitudes $E_{\rm A} (x_{\rm A},t_{\rm A})$ seen by Alice at positions $x_{\rm A}$ at times $t_{\rm A}$ equal
	\begin{eqnarray} \label{1A}
		E_{\rm A} (x_{\rm A},t_{\rm A}) &=& E_{\rm A} (x_{\rm A} - sc t_{\rm A},0)
	\end{eqnarray}
	where the initial electric field amplitudes of the wave packet are given by $E_{\rm A} (x_{\rm A},0)$. Here $s=-1$ and $s=1$ correspond to wave packets propagating in the direction of decreasing and increasing $x_{\rm A}$ respectively. Hence, if the physical properties of a wave packet seen by Alice are known at one instant in time, they are known at all times. The same applies to the electric field amplitudes $E_{\rm B} (x_{\rm B},t_{\rm B})$ seen  at $(x_{\rm B},t_{\rm B})$ by a second observer\textemdash called Bob (B) \textemdash who is travelling at a constant velocity $v_{\rm B}$ relative to Alice, wherefore 
	\begin{eqnarray} \label{1B}
		E_{\rm B} (x_{\rm B},t_{\rm B}) &=& E_{\rm B} (x_{\rm B} - sc t_{\rm B},0) 
	\end{eqnarray}
	in analogy to Eq.~(\ref{1A}). Hence, the electric field amplitudes perceived by both Alice and Bob at any position and time are only characterised by the values of the parameters $\chi_i = x_i -sct_i$ with $i= {\rm A},{\rm B}$. In the remainder of this paper, we shall use a shorthand notation and replace $E_i (x_i,t_i) $ by $E_i(\chi_i)$.  
	
	The principle of relativity also suggests that the electric and magnetic field transformations from observer A to observer B and vice versa need to be formally the same. The only difference is that the relative speed of their reference frames changes from $v_{\rm B}$ to $v_{\rm A} = - v_{\rm B}$. This suggests a linear dependence between electric field amplitudes $E_{\rm B} (\chi_{\rm B})$ and $E_{\rm A} (\chi_{\rm A})$ at the same point in the spacetime diagram since this transformation is the only transformation that remains formally the same when reversed. We therefore assume in the following that
	\begin{eqnarray} \label{2}
		E_{\rm B} (\chi_{\rm B}) &=& \xi_{\rm BA} \, E_{\rm A} (\chi_{\rm A}) 
	\end{eqnarray}
	where the coordinates $\chi_{\rm A}$ and $\chi_{\rm B}$ specify the same spacetime trajectory and $\xi_{\rm BA}$ denotes a transformation constant. Analogously, we also know that
	\begin{eqnarray} \label{3}
		E_{\rm A} (\chi_{\rm A}) &=& \xi_{\rm AB} \, E_{\rm B} (\chi_{\rm B}). 
	\end{eqnarray}
	Furthermore, the principle of relativity tells us that the transformation constants $\xi_{\rm AB}$ and $\xi_{\rm BA}$ relate to each other such that
	\begin{eqnarray} \label{4}
		\xi_{\rm AB}(s,v_{\rm B}) &=& \xi_{\rm BA}(s,-v_{\rm B}) \, ,
	\end{eqnarray}
	since the direction of propagation $s$ of the wave packet is the same in both reference frames, but the relative speed of the frames changes sign. When combining Eqs.~(\ref{2})--(\ref{4}) we therefore find that
	\begin{eqnarray} \label{5}
		\xi_{\rm BA}(s,v_{\rm B}) = 1/\xi_{\rm BA}(s,-v_{\rm B}) \, .   
	\end{eqnarray}
	In the following, this is taken into account when we determine $ \xi_{\rm AB}$ and  $\xi_{\rm BA}$. Whilst some quantisations based on the vector potential require a gauge fixing condition, which may not be relativistically invariant, in this paper we shall deal directly with the gauge invariant electric and magnetic field observables.  As a consequence, Eqs.~(\ref{2}) and (\ref{3}) are the only transformation conditions required in this paper. 
	
	Next we notice that the spatial and time translational symmetries of the EM field tell us that the above relations must hold for all spacetime coordinates $\chi_{\rm A}$ and $\chi_{\rm B}$. Hence the transformation factors $\xi_{\rm BA}$ and $\xi_{\rm AB}$ can depend on the direction of propagation $s$ and on the relative speed $v_{\rm B}$ of Bob's reference frame with respect to Alice's, but not on where and when the electric and magnetic field amplitudes are measured. The above arguments thus reduce the question, how do local electric and magnetic field observables transform from one inertial frame to another, to the simpler question, how do they transform at a single point in the spacetime diagram? Nevertheless, the above equations are not enough to determine the transformation factor $\xi_{\rm BA}$ in Eq.~(\ref{2}). To answer this question, an additional assumption is needed.
	
	Our final assumption in the following derivation of the relativistic Doppler effect is based on energy conservation. To implement this we consider a certain ``box" which is a volume of spacetime points obtained by identifying a finite-sized interval along the $x$ axis and extending it to also contain all future and past points along light-like trajectories passing through this interval. By integrating over the positions inside the ``box" at a fixed time we can calculate the total amount of energy that it contains. By construction, the same ``box" must contain the same amount of energy in Alice's and Bob's reference frames since it contains the same physical system in both cases. Nevertheless, as Alice and Bob experience space and time differently, the same ``box" appears to have a different size in each of their frames. For example, parts of the wave packet that occur simultaneously in the frame of observer A appear at different times in the reference frame of observer B. In addition, the density of the possible trajectories of light changes when moving from one inertial frame to another; however, the total number of world-lines in the ``box" must remain the same. Taking this into account, we can finally identify the dependence of $\xi_{\rm BA}$ on $s$ and on $v_{\rm B}$. When applying Fourier transforms to local electric field amplitudes, we obtain the usual frequency, wavelength and amplitude changes of the relativistic Doppler effect.
	
	\begin{figure*}[t]
		\centering
		\includegraphics[width = 1.6\columnwidth]{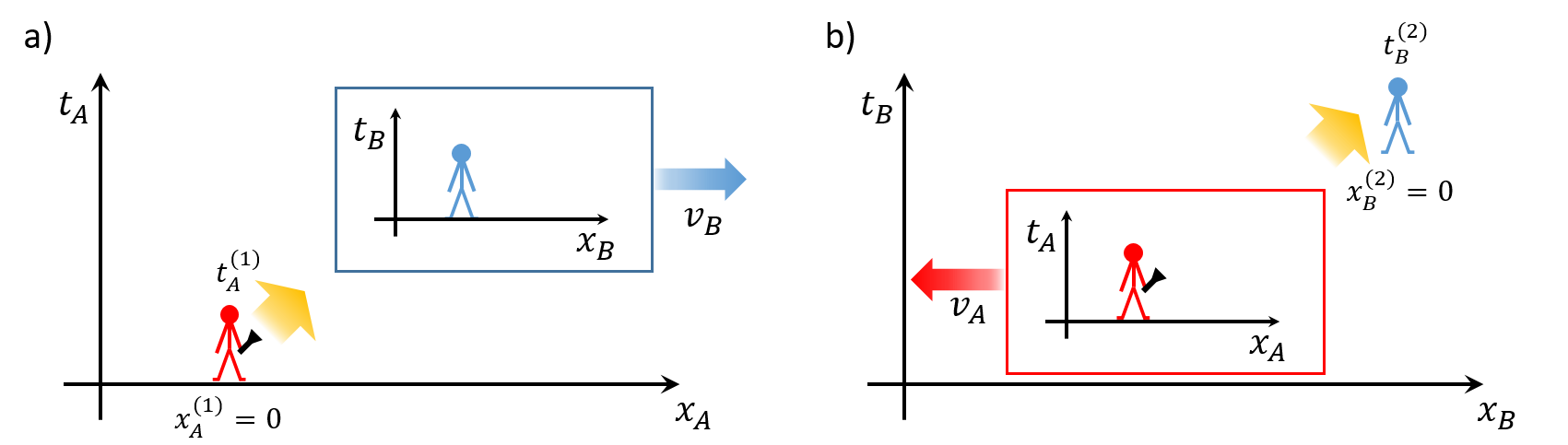}
		\caption{Schematic view of two observers, Alice (a) and Bob (b), in different inertial reference frames which move with respect to each other at constant velocity. For simplicity, we assume here that both observers are based at the origin of their respective coordinate system and share the same position at an initial time $t_{\rm A} = t_{\rm B} = 0$. Suppose Alice emits a short light pulse from her position towards Bob at a time when her clock reads $t_{\rm A}^{(1)}$ which Bob then receives when his clock reads $t_{\rm B}^{(2)}$. By comparing these two times, the ratio of their spacetime coordinates, i.e.~$\chi_{\rm B}/\chi_{\rm A}$, can be determined.}
		\label{Fig:experiment}
	\end{figure*}
	
	This paper is structured as follows. Section \ref{Sec:2} reviews the relativistic Doppler effect in classical physics. We first study how the coordinates $\chi_{\rm A}$ and $\chi_{\rm B}$ of two inertial observers A and B relate to each other when they correspond to the same point in the spacetime diagram. Afterwards, we derive the transformation factors $\xi_{\rm BA}$ and $\xi_{\rm AB}$ in Eqs.~(\ref{2}) and (\ref{3}) by imposing the above described conditions. In Section \ref{Sec:3}, we use a local photon approach and proceed as described in Refs.~\cite{Ali,Southall:2019zdm,Hodgson:2021eyb} to quantise the EM field in different inertial reference frames. Section \ref{Sec:4} combines this description with the results of Section \ref{Sec:2} to obtain a quantum picture of the relativistic Doppler effect. Given the principle of relativity, neither observer A nor observer B should be able to perform measurements on photonic wave packets which tell them about their absolute speed. Taking this into account, we find that the local photon annihilation operators of Alice and Bob are the same when they refer to the same location in the spacetime diagram. However, the transformations of the annihilation operators of monochromatic photons are more complex. Finally, we summarise our findings in Section \ref{Sec:5}.
	
	\section{The relativistic Doppler effect}
	\label{Sec:2}
	
	The motion of an observer affects both the time and the distance separating two events in spacetime. In the case of two observers Alice (A) and Bob (B) in flat 1+1 dimensional spacetime (Minkowski space), this difference in duration and separation can be expressed as a transformation between their natural coordinates $\chi_{\rm A}$ and $\chi_{\rm B}$. In this section, we provide a derivation of the transformation between the coordinates of an observer at rest and an observer moving with a constant velocity. Afterwards, we use this to identify the constant $\xi_{\rm BA}$ for the electric field amplitude transformation in Eq.~(\ref{2}).
	
	\subsection{Coordinate transformations}
	
	Suppose our first observer, Alice, is at rest and provides a point of reference whilst our second observer, Bob, travels at a constant velocity $v_{\rm B}$ relative to Alice along the $x$ axis. As illustrated in Fig.~\ref{Fig:experiment}, the position and time at which an event takes place from Alice's point of view are denoted $x_{\rm A}$ and $t_{\rm A}$ respectively. Analogously, from Bob's point of view, events take place at positions $x_{\rm B}$ and times $t_{\rm B}$. For simplicity, we assume in the following that both observers, who are stationary with respect to their own coordinate systems, are located at the origin. This means that Alice's position is given by $x_{\rm A} = 0$ while Bob's position is given by $x_{\rm B} = 0$ for all times $t_{\rm A}$ and $t_{\rm B}$ respectively. Moreover, we assume that Bob meets Alice only once at an initial time $t_{\rm A} = t_{\rm B} = 0$.
	
	In the stationary reference frame, light with a well-defined direction of propagation $s$ travels along the $x_{\rm A}$ axis at the speed of light $c$. Therefore, if Alice observes any localised pulse of light, its position $x_{\rm A}$ at any time $t_{\rm A}$ satisfies the relation
	\begin{eqnarray}
		\label{Alice1}
		\chi_{\rm A} = x_{\rm A} - sc t_{\rm A} = \text{const.}
	\end{eqnarray}
	Here $\chi_{\rm A}$ coincides with the position $x_{\rm A}$ of the light pulse at $t_{\rm A} = 0$. The speed of light measured relative to the rest frame of an inertial observer is always constant and independent of the motion of the source. Hence, from Bob's point of view, the position $x_{\rm B}$ of the same light pulse at any time $t_{\rm B}$ satisfies the relation
	\begin{eqnarray}
		\label{Bob1}
		\chi_{\rm B} = x_{\rm B} - sc t_{\rm B} = \text{const.}
	\end{eqnarray}
	In general $\chi_{\rm B}$ does not equal $\chi_{\rm A}$, but the direction of propagation $s$ must be the same in both reference frames. As both Eqs.~(\ref{Alice1}) and (\ref{Bob1}) must be satisfied, we have
	\begin{eqnarray}
		\label{relation1}
		\chi_{\rm B} &=& \kappa \, \chi_{\rm A} \, .
	\end{eqnarray}
	The constant $\kappa$ provides a connection between the coordinate $\chi_{\rm A}$ adopted by Alice and the coordinate $\chi_{\rm B}$ adopted by Bob, thereby establishing a relation between the coordinates of identical world-lines. Considering the cases where $s=-1$ and $s=+1$ and solving the above equations, one can derive the point-like coordinate transformations between spacetime coordinates $(x_{\rm A},t_{\rm A})$ and $(x_{\rm B},t_{\rm B})$ which refer to the same point in the spacetime diagrams of Alice and Bob.
	
	\begin{figure*}[t]
		\centering
		\includegraphics[width=1.4 \columnwidth]{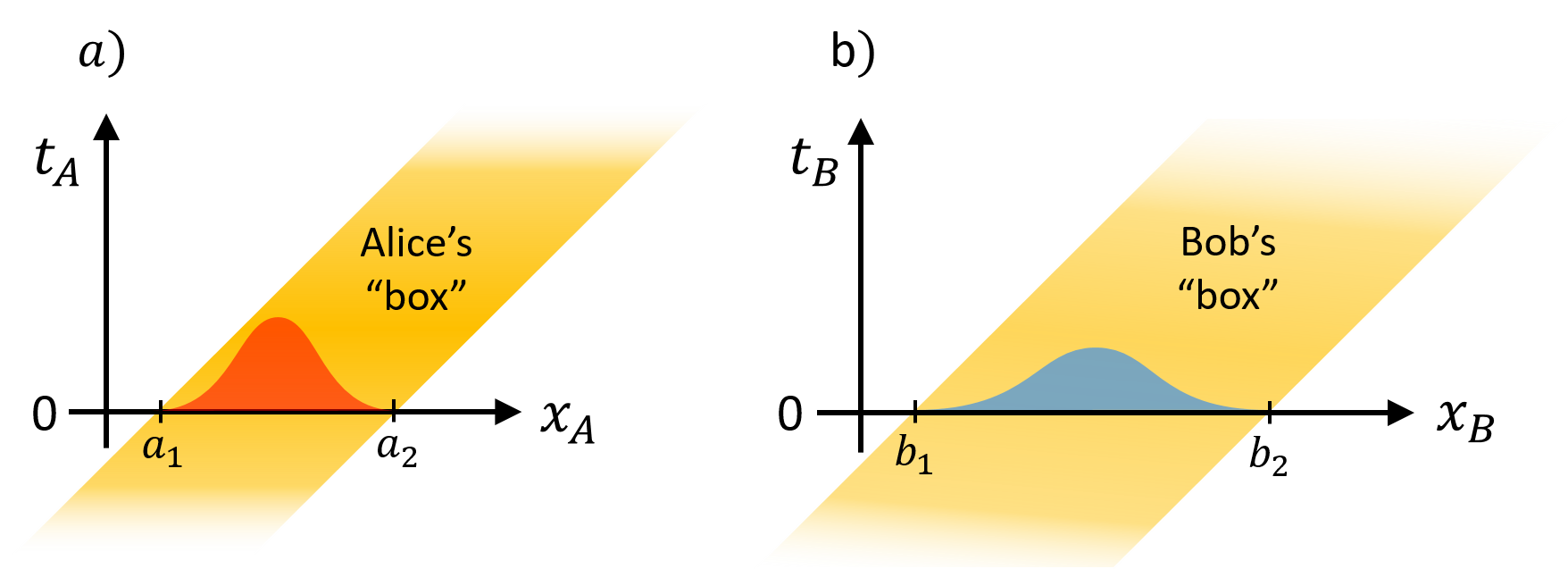}
		\caption{The figures show the spacetime diagram of a short light pulse permanently confined to a spacetime ``box" from both Alice's (a) and Bob's (b) points of view.  From Alice's point of view the ``box" extends from $x_{\rm A} = a_1$ to $x_{\rm A} = a_2$.  The amplitude of the light pulse in Alice's frame is illustrated by the red wave form and remains within the box for all $t_{\rm A}$.  From Bob's point of view the same ``box" extends from $x_{\rm B} = b_1$ to $x_{\rm B} = b_2$. In Bob's frame the width of the ``box" is increased relative to Alice by a factor of $\gamma(1+s\beta)$ where $\beta = v_{\rm B}/c$. The wave form seen by Bob is shown in blue and remains in the ``box" for all $t_{\rm B}$.}
		\label{Fig:energyflux}
	\end{figure*}
	
	To determine the relating constant $\kappa$ in Eq.~(\ref{relation1}) we assume that Alice sends a short light pulse from her own position at $x_{\rm A}^{(1)} = 0$ at a time $t_{\rm A}^{(1)}$ to Bob (cf.~Fig.~\ref{Fig:experiment}). From Bob's point of view, the light is emitted from a position $x_{\rm B}^{(1)}$ at a time $t_{\rm B}^{(1)}$ and arrives at his position $x_{\rm B}^{(2)} = 0$ when his watch reads a time $t_{\rm B}^{(2)}$. As Alice and Bob have both been placed at the origin of their respective coordinate systems, Eq.~(\ref{relation1}) tells us that
	\begin{eqnarray}
		\label{relation2}
		t_{\rm B}^{(2)} &=& \kappa \, t_{\rm A}^{(1)} \, .
	\end{eqnarray}
	According to Alice, a right-moving light pulse is received by Bob at a position $x_{\rm A}^{(2)}$ at a time $t_{\rm A}^{(2)}$ where $t^{(2)}_{\rm A} = t^{(1)}_{\rm A} + x_{\rm A}^{(2)}/c$ and $x_{\rm A}^{(2)} = v_{\rm B}t_{\rm A}^{(2)}$. Putting these two relations together we find that 
	\begin{eqnarray}
		\label{delay1}
		t^{(1)}_{\rm A} &=& (1-\beta) \, t^{(2)}_{\rm A}
	\end{eqnarray}
	where the constant $\beta$ is defined as 
	\begin{eqnarray}
		\beta &=& v_{\rm B}/c \, .
	\end{eqnarray} 
	Analogously, one can also show that
	\begin{eqnarray}
		\label{delay2}
		t_{\rm B}^{(2)} &=& (1+\beta) \, t_{\rm B}^{(1)} \, .
	\end{eqnarray}
	Eqs.~(\ref{delay1}) and (\ref{delay2}) specify the relationship between the times at which the light is emitted and received from the points of view of both observers.
	
	In general, the time elapsed between two events in Alice's reference frame is different to the time elapsed between the same two events in Bob's frame \cite{timedilation,Crouse:2018slq, GWINNER2005,Saathoff2003}. The reason for this is that a moving clock ticks at a different rate than a stationary clock \cite{Hafele,Unnikrishnan:2004nj}. To properly take this into account, as the position at which Bob receives the signal is moving relative to Alice, we suppose that
	\begin{eqnarray}
		\label{gamma1}
		t^{(2)}_{\rm A} = \gamma \, t^{(2)}_{\rm B} \, .
	\end{eqnarray} 
	Furthermore, as the position of the emitter is stationary with respect to Alice, but now moving in the opposite direction with respect to Bob, it must also be that
	\begin{eqnarray}
		\label{gamma2}
		t^{(1)}_{\rm B} = \gamma \, t^{(1)}_{\rm A} \, .
	\end{eqnarray}
	By combining Eqs.~(\ref{delay1})-(\ref{gamma2}) it can now be shown that
	\begin{eqnarray}
		\label{gamma3}
		\gamma = 1/\sqrt{1-\beta^2} \, .
	\end{eqnarray}
	Since $\gamma$ is always larger than 1, clocks run slower in a moving frame. By putting together Eqs.~(\ref{delay2}) and (\ref{gamma2}), we moreover find that $\kappa = \gamma(1+\beta)$. By carrying out a similar calculation for left-propagating light we eventually obtain the complete relation
	\begin{eqnarray}
		\label{trans1}
		\chi_{\rm B} &=& \gamma(1+s\beta) \, \chi_{\rm A} \, .
	\end{eqnarray}
	As one would expect, the constant $\kappa$ in Eq.~(\ref{relation2}) depends on both the direction of propagation of light $s$ and the relative velocity $v_{\rm B}$ between Alice and Bob.
	
	\subsection{Field amplitude transformations}
	\label{Sec:2B}
	
	Although light propagates at the same speed in all reference frames, depending upon the direction of the wave packet and the relative speed between Alice and Bob, wave packets will appear to be either stretched or squeezed from the point of view of a moving observer. This occurs due to the difference in how space and time is perceived by a moving observer compared to an observer at rest, as we described in the previous subsection. As a consequence of this transformation, a wave packet of light may appear to have different energies in Alice's and Bob's reference frames. However, the energy associated with the same number of light trajectories must be the same for Alice and Bob simultaneously. The purpose of this subsection is to exploit energy conservation in order to determine the transformation coefficient $\xi_{\rm BA}$ defined in Eq.~(\ref{2}). To do so, we now investigate the energy of a bundle of light trajectories from both Alice's and Bob's points of view.  
	
	More concretely, as illustrated in Fig.~\ref{Fig:energyflux}, we now place a ``box" that confines a bundle of light-like world-lines with coordinates $\chi_{\rm A} \in [a_1,a_2]$ into the spacetime diagram of Alice. From Alice's point of view, the ``box" has a width $\Delta x_{\rm A} = a_2 - a_1$, which corresponds to an instant in time, and contains all past and future points associated with this spatial interval. Entirely analogous to Alice, the same bundle of world-lines creates an analogous ``box" in Bob's spacetime diagram, but its width $\Delta x_{\rm B}$ along the $x_{\rm B}$ axis differs from $\Delta x_{\rm A}$. As shown in Fig.~\ref{Fig:energyflux}, we denote the endpoints of the ``box" in Bob's spacetime diagram at $t_{\rm B} = 0$ by $b_1$ and $b_2$. Hence $\Delta x_{\rm B} = b_2 - b_1$ and Bob's world-line coordinates are $\chi_{\rm B} \in [b_1,b_2]$.
	
	Suppose now that $h_{\rm A}$ and $h_{\rm B}$ denote the density of the bundle of world-lines in Alice's and in Bob's reference frame respectively. By taking into account conservation of the total number of world-lines within the ``box," we see that
	\begin{eqnarray}
		\label{conservation1}
		h_{\rm A} \, \Delta x_{\rm A} &=& h_{\rm B} \, \Delta x_{\rm B} \, . 
	\end{eqnarray}
Since $\beta$ and $\gamma$ are defined such that
\begin{eqnarray}
	\label{extra}
	\left[ \gamma (1+ s\beta)\right]^{-1} &=& \gamma (1- s\beta) \, ,
\end{eqnarray} 
	by using Eq.~(\ref{trans1}) we can show that
	\begin{eqnarray}
		\label{conservation2}
		h_{\rm B} &=& \gamma (1-s\beta) \, h_{\rm A} \, .
	\end{eqnarray}
	Due to the change in the space and time coordinates between Alice's and Bob's reference frames, from Bob's point of view, the position density of the trajectories of light has changed by a factor of $\gamma(1-s\beta)$.
	
	The space and time transformations that take place between Alice's and Bob's reference frames are unknown to them. We, however, knowing that there is a change in the energy density along the $\chi$ axis must take this into account when we look more closely at Alice's and Bob's energy observables. Doing so, we conclude that the actual amounts of energy associated with a fixed number of trajectories in both Alice's and Bob's ``boxes" are only the same when \cite{Bennett}
	\begin{eqnarray}
		\label{fieldenergy1}
		&& \hspace*{-1cm} \frac{A\varepsilon}{2 h_{\rm A}} \int_{a_1}^{a_2} \text{d}\chi_{\rm A} \left[ E_{\rm A}(\chi_{\rm A})^2 + c^2 \, B_{\rm A}(\chi_{\rm A})^2 \right]  \notag \\
		&=& \frac{A\varepsilon}{2 h_{\rm B}} \int_{b_1}^{b_2} \text{d}\chi_{\rm B} \left[ E_{\rm B}(\chi_{\rm B})^2 + c^2 \, B_{\rm B}(\chi_{\rm B})^2 \right] \, .
	\end{eqnarray}
	Here $A$ is the area occupied by the EM field in the $y-z$ plane, which is unchanged by boosts along the $x$ axis, and $\varepsilon$ is the permittivity of free space. Eq.~(\ref{fieldenergy1}) can now be used to obtain an expression for $\xi_{\rm BA}$. As the ratio between the electric and magnetic field amplitudes of travelling waves must be the same in every reference frame, which implies that $B_{\rm B} (\chi_{\rm B}) = \xi_{\rm BA} \, B_{\rm A} (\chi_{\rm A}) $ in analogy to Eq.~(\ref{2}), and substituting Eqs.~(\ref{2}), (\ref{trans1}) and (\ref{conservation2}) into Eq.~(\ref{fieldenergy1}), we conclude that
	\begin{eqnarray}
		\label{finaltrans1}
		\gamma(1+s\beta) \, \xi_{\rm BA}^2 &=& \gamma(1-s\beta)
	\end{eqnarray}
	and hence
	\begin{eqnarray}
		\label{fieldtransforms8}
		\xi_{\rm BA}(s, v_{\rm B}) &=& \gamma(1-s\beta) \, .
	\end{eqnarray}
	This expression for the electric field transformation constant $\xi_{\rm BA}$ satisfies Eq.~(\ref{5}). 
	
	For completeness, let us add that the usual expressions for the total energy of the EM field in Alice's and Bob's reference frames are given by \cite{Bennett}
	\begin{eqnarray}
		\label{fieldenergy11}
		H^{(i)}_{\rm energy} &=& \frac{A\varepsilon}{2}\int_{-\infty}^{\infty}\text{d}\chi_i \left[ E_i(\chi_i)^2 + c^2 \, B_i(\chi_i)^2 \right] ~~
	\end{eqnarray}
	with $i = {\rm A},{\rm B}$. Not taking into account the different densities $h_{\rm A}$ and $h_{\rm B}$ of light trajectories, but nevertheless employing the results in Eqs.~(\ref{trans1}) and (\ref{fieldtransforms8}), we find that the above expressions are related by the relativistic transformation
	\begin{eqnarray}
		\label{fieldenergy22}
		H^{({\rm B})}_{\rm energy} &=& \gamma\left(1-s\beta\right)H^{({\rm A})}_{\rm energy} \, .
	\end{eqnarray}
	There now seems to be a difference in the energy of the EM field seen by Alice and the energy seen by Bob. However, as we have seen above, energy conservation is restored when we correctly account for the space and time transformations between Alice's and Bob's reference frames. The above discussion also shows that the total amount of energy of the EM field in a given reference frame $i$ can only be calculated up to an overall factor unless there is a way of determining its world-line density $h_i$. 
	
	\subsection{Frequency and wavelength transformations}
	\label{Sec:Doppler1}
	
	Whilst the previous subsection deals only with changes in the magnitude of the local energy, the Doppler effect is normally associated with frequency and wavelength shifts of monochromatic waves seen by two different inertial observers \cite{Michel2014}. These shifts are not surprising since the frequency of a monochromatic wave seen by Alice and Bob is the number of complete wavelengths that pass their positions per unit time. Frequency and wavelength are therefore strongly connected with the clock or meter being used as a measuring device \cite{Wilmshurst}. For completeness, we therefore now also have a closer look at the momentum representation of electric field amplitudes. Since the complex electric field amplitudes $\widetilde{E}_i(k_i,t_i)$ and $E_i(x_i,t_i)$ in momentum and in position space relate to each other via a Fourier transform, we have \cite{Hodgson:2021eyb}
	\begin{eqnarray}
		\label{frequency1new}
		E_i(x_i,t_i) &=& {1 \over \sqrt{2\pi}} \int_{-\infty}^{\infty} \text{d}k_i \, {\rm e}^{{\rm i} s k_i x_i} \, \widetilde{E}_i(k_i,t_i) 
	\end{eqnarray} 
	with
	\begin{eqnarray}
		\label{frequency1new2}
		\widetilde{E}_i(k_i,t_i) &=& {\rm e}^{-{\rm i}ck_i t_i} \, \widetilde{E}_i(k_i,0)
	\end{eqnarray} 
	for $i={\rm A},{\rm B}$. Here $ck_{\rm A}$ and $ck_{\rm B}$ are the frequencies of a monochromatic wave observed by Alice and Bob respectively. 
	
	By again using the coordinates $\chi_i$ defined in Eqs.~(\ref{Alice1}) and (\ref{Bob1}), and taking into account the main result of the previous subsection, i.e.~by combining Eqs.~(\ref{2}) and (\ref{fieldtransforms8}), we see that
	\begin{eqnarray} \label{2new}
		E_{\rm B} (\chi_{\rm B}) &=& \gamma(1-s\beta) \, E_{\rm A} (\chi_{\rm A}) \, .
	\end{eqnarray}
	Substituting the above Fourier transform into this relation yields
	\begin{eqnarray}
		\label{frequency1}
		&&\hspace*{-1cm} {1 \over \sqrt{2\pi}} \int_{-\infty}^{\infty} \text{d}k_{\rm B} \, {\rm e}^{{\rm i}sk_{\rm B} \chi_{\rm B}} \, \widetilde{E}_{\rm B} (k_{\rm B},0) \notag \\
		&=& {1 \over \sqrt{2\pi}} \, \gamma(1-s\beta)\,\int_{-\infty}^{\infty} \text{d}k_{\rm A} \, {\rm e}^{{\rm i}sk_{\rm A} \chi_{\rm A}} \, \widetilde{E}_{\rm A}(k_{\rm A},0) \, . ~~
	\end{eqnarray} 
	By taking the inverse Fourier transformation of both sides of Eq.~(\ref{frequency1}) with respect to the coordinate $\chi_{\rm B}$, we can now show that
	\begin{eqnarray}
		\label{frequency2}
		\widetilde{E}_{\rm B}(k_{\rm B},0) &=& {1 \over 2\pi} \, \gamma(1-s\beta) \int_{-\infty}^{\infty} \text{d} \chi_{\rm B} \int_{-\infty}^{\infty} \text{d} k_{\rm A} \notag \\
		&& \times {\rm e}^{{\rm i}s(k_{\rm A} \chi_{\rm A}- k_{\rm B} \chi_{\rm B})} \, \widetilde{E}_{\rm A}(k_{\rm A},0) \, . 
	\end{eqnarray} 
	After substituting $\chi_{\rm A}$ for $\chi_{\rm B}$ using Eq.~(\ref{trans1}), the $\chi_{\rm B}$ integration can be solved.  This integration leads us to the final result
	\begin{eqnarray}
		\label{frequency3}
		\widetilde{E}_{\rm B}(k_{\rm B},0) &=& \widetilde{E}_{\rm A}(\gamma(1+s\beta)k_{\rm B},0) \, .
	\end{eqnarray} 
	The above equality specifies the relationship between the Fourier components $\widetilde{E}_{\rm A}(k_{\rm A},0)$ and $\widetilde{E}_{\rm B}(k_{\rm B},0) $ of the electric field amplitudes measured by Alice and by Bob. If, for instance, $\widetilde{E}_{\rm A}(k_{\rm A},0)$ is non-zero for a wave with a single frequency $c k_{\rm A}$ only, then Bob observes a monochromatic wave with frequency 
	\begin{eqnarray}
		\label{frequency4}
		c k_{\rm B} &=& \gamma(1-s\beta) \, c k_{\rm A} \, .
	\end{eqnarray} 
	This shift in frequency is consistent with previous derivations of the relativistic Doppler shift for light propagating in the $s$ direction \cite{French, Longhurst}.
	
	\section{The quantised EM field in the stationary frame}
	\label{Sec:3}
	
	For a long time it has been believed that photons do not have a wave function and that light cannot be localised \cite{Reeh,Bialynicki-Birula,Sipe}. However, quantum  physics should apply to all particles and photons should not be an exception. For example, when a single-photon detector clicks, it measures the position of the arriving photon at that instant in time \cite{Kuhn2002,Kuhn2012}.  Defining a time of arrival operator for a localised photon detector, however, has been a significant problem and could not be achieved within the standard Hilbert space of the quantised EM field \cite{Allcock1, Allcock2, Aharonov, Delgado, Schlichtinger}. The origin of the wave function problem was that many authors like to identify the wave function of the photon with its electric field amplitudes, but the complex electric field amplitudes at different positions do not commute. The eigenstates of the electric field observable are therefore not local, although they can be made to appear local by altering the scalar product that is used to calculate the overlap of quantum state vectors \cite{Ali,Hawton}. 
	
	An alternative way of establishing the wave function of a single photon is to double the Hilbert space of the quantised EM field to include both positive and negative frequency photons and to separate light from its carriers \cite{Southall:2019zdm,Hodgson:2021eyb,Hodgson2}. The carriers of the quantised EM field in momentum space are non-local monochromatic waves. The Fourier transforms of these carriers, however, so-called {\em blips} (bosons localised in position) form a complete set of pairwise orthonormal local carriers of the quantised EM field in position space. Similar to how a point mass is a carrier for a gravitational field, blips are carriers of non-local electric and magnetic field amplitudes. When expressing the observables of the electric and magnetic field in free space in terms of blip annihilation and creation operators, these include contributions from blips at all points along the position axis. By applying a constraint to the blip dynamics, a relativistically form-invariant representation of the EM field is derived. Below these expressions are used to derive a transformation between blips in Alice's and Bob's reference frames.         
	
	\subsection{Local photons}
	
	Let us first have a closer look at the modeling of the quantised EM field in Alice's resting reference frame. Here blips are characterised by their position $x_{\rm A} \in (-\infty, \infty)$ at a given time $t_{\rm A} \in (-\infty, \infty)$ as well as their direction of propagation $s $ and their polarisation $\lambda$.  For boosts and translations along the $x_{\rm A}$ and $t_{\rm A}$ axes, $s$ and $\lambda$ are invariant.  The parameter $s = \pm1$ denotes propagation in the direction of increasing and decreasing $x_{\rm A}$ respectively.  We shall assume that $\lambda = \mathsf{H}, \mathsf{V}$ are two linear polarisations orthogonal to the $x_{\rm A}$ axis \cite{Hodgson:2021eyb,Southall:2019zdm}. The creation operator $a^\dagger_{s\lambda}(x_{\rm A})$ adds to the system a single blip located at a position $x_{\rm A}$ at a time $t_{\rm A} =0$ with direction of propagation $s$ and polarisation $\lambda$. In the above $\dagger$ denotes complex conjugation and distinguishes $a^\dagger_{s\lambda}(x_{\rm A})$ from the annihilation operator $a_{s\lambda}(x_{\rm A})$ which removes the same blip from the system. 
	
	For consistency with Maxwell's equations, all blips must propagate at the speed of light. This constraint imposes the following condition: at some time $t_{\rm A}$, the time-evolved operator $U_{\rm A}(t_{\rm A},0) \, a^\dagger_{s\lambda}(x_{\rm A}) \, U^\dagger_{\rm A}(t_{\rm A},0)$ must be equivalent to the blip creation operator at a position $x_{\rm A}-sct_{\rm A}$. Hence
	\begin{eqnarray}
		\label{dyn}
		U_{\rm A}(t_{\rm A},0) \, a^\dagger_{s\lambda}(x_{\rm A}) \, U^\dagger_{\rm A}(t_{\rm A},0) &=& a^\dagger_{s\lambda}(x_{\rm A} -sct_{\rm A}) ~~
	\end{eqnarray}
	where $U_{\rm A}(t_{\rm A},0)$ is the time evolution operator of the quantised EM field in Alice's reference frame. As a consequence of this, all blips characterised by the same coordinate $\chi_{\rm A} = x_{\rm A}-sct_{\rm A}$ are identical.  From this point onwards we shall therefore denote blip creation and annihilation operators in Alice's reference frame $a^\dagger_{s\lambda}(\chi_{\rm A})$ and $a_{s\lambda}(\chi_{\rm A})$ respectively. Blips that are characterised by non-identical values of $\chi_{\rm A}$, $s$ or $\lambda$ are distinguishable from one another, and therefore pairwise orthogonal. Hence, we can determine that
	\begin{eqnarray}
		\label{Comm1}
		\left[a_{s\lambda}(\chi_{\rm A}), a^\dagger_{s'\lambda'}(\chi'_{\rm A})\right] &=& \delta_{ss'}\,\delta_{\lambda\lambda'}\,\delta(\chi_{\rm A}-\chi'_{\rm A}) \, .
	\end{eqnarray} 
	All creation operators commute with one another, as do the annihilation operators.
	
	\subsection{Field observables in position representation}
	
	\begin{figure}[t]
		\centering
		\includegraphics[width= 0.9\columnwidth]{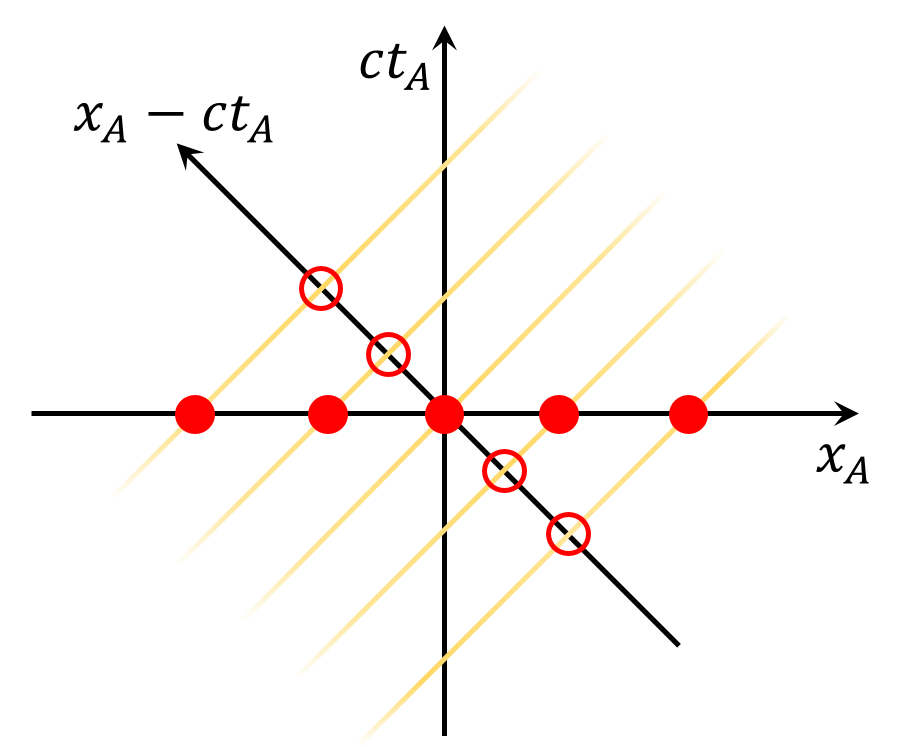}
		\caption{The diagram shows the contribution of right-propagating blips to the field observables measured by Alice at the origin.  From one point of view, blips distributed along the $x_{\rm A}$ axis (solid red) contribute non-locally to the field observable.  As blips at one point in spacetime can be identified with blips at all other points along their world-lines (marked in yellow), blips distributed along $x_{\rm A}-ct_{\rm A}$ axis (hollow red) provide an equivalent contribution to the field observables as blips along the $x_{\rm A}$ axis on the same world-line.}
		\label{Fig:timeA}
	\end{figure} 
	
	As mentioned already above, Alice's electric and magnetic field observables $E_{\rm A}(x_{\rm A},t_{\rm A})$ and $B_{\rm A}(x_{\rm A},t_{\rm A})$ in the Heisenberg picture, which are measured at a position $x_{\rm A}$ and time $t_{\rm A}$, contain non-local contributions from blips at all points along the $x_{\rm A}$ axis \cite{Hodgson:2021eyb, Casimir}. All her blips contribute simultaneously to these field observables independently of their separation from Alice.  For a moving observer like Bob, however, events that take place along the $x_{\rm A}$ axis at a fixed time $t_{\rm A}$ are no longer simultaneous and do not occur at a single time $t_{\rm B}$. For this reason, one has to be careful when defining field observables by superposing simultaneous blips at different positions along the $x$ axis. Fortunately, blips can be identified with all other blips along their individual trajectories which allows us to represent the field observables as a non-local superposition over the $\chi_{\rm A} = x_{\rm A}-sct_{\rm A}$ coordinates. This change in representation is illustrated in Fig.~\ref{Fig:timeA} for a field propagating to the right.  In this diagram the field observable at the origin can be determined as a non-local superposition of blips along the $x_{\rm A}$ axis at a fixed $t_{\rm A}$ (marked in red).  Alternatively, they can be identically determined as a superposition of blips dispersed along the $\chi_{\rm A}$ axis.  By fixing the time $t_{\rm A}$, this new representation returns to it's original form.
	
	As in Section \ref{Sec:2} and for simplicity, we shall restrict ourselves in the following to only one polarisation $\lambda$. Let us say $\lambda = {\sf H}$. In this case, the electric field is horizontally polarised whereas the magnetic field is vertically polarised, but we consider only the amplitude of Alice's electric and magnetic field vectors. Taking this into account, $E_{\rm A}(\chi_{\rm A})$ and $B_{\rm A}(\chi_{\rm A})$ can be written as
	\begin{eqnarray}
		\label{Fields1}
		E_{\rm A}(\chi_{\rm A}) &=& \sum_{s=\pm1} \int_{-\infty}^{\infty}\text{d} \chi'_{\rm A} \, c\, \mathcal{R}(\chi_{\rm A}-\chi'_{\rm A}) \, a_{s\mathsf{H}}(\chi'_{\rm A}) \notag \\
		&& + {\rm H.c.} \, , \notag \\
		B_{\rm A}(\chi_{\rm A}) &=& \sum_{s=\pm1} \int_{-\infty}^{\infty}\text{d} \chi'_{\rm A}\, s \,\mathcal{R}(\chi_{\rm A}-\chi'_{\rm A}) \, a_{s\mathsf{H}}(\chi'_{\rm A}) \notag \\
		&& + {\rm H.c.} 
	\end{eqnarray} 
	These operators represent the observables of the electric and magnetic field amplitudes respectively at position $x_{\rm A} = \chi_{\rm A}$ at $t_{\rm A}=0$ and at every position along the $\chi_{\rm A} = \text{contant}$ trajectory. In the expressions above, the contribution of each blip to Alice's field observables is weighted by a non-local distribution $\mathcal{R}(\chi_{\rm A}-\chi'_{\rm A})$, which we shall refer to as the regularisation function. By taking into account that a single monochromatic photon has the positive energy $\hbar c |k_{\rm A}|$, the function $\mathcal{R}(\chi_{\rm A}-\chi'_{\rm A})$ can be determined explicitly and can be shown to be given by \cite{Ali,Hodgson:2021eyb,Hodgson2}
	\begin{eqnarray}
		\label{Reg1}
		\mathcal{R}(\chi_{\rm A}-\chi'_{\rm A}) &=& -\sqrt{\frac{\hbar}{4\pi\varepsilon cA}}\cdot\frac{1}{|\chi_{\rm A}-\chi'_{\rm A}|^{3/2}} \, .
	\end{eqnarray}
	This highly non-local function is closely related to the 1-dimensional Feynman propagator for two excitations of the EM field observables.  The Feynman propagator is non-local whereas correlations between blips are strictly localised. As $\mathcal{R}(\chi_{\rm A}-\chi'_{\rm A})$ is non-zero for all values of $\chi_{\rm A} \neq \chi'_{\rm A}$, we view each blip as carrying with it static and non-local electric and magnetic fields. 
	
	\subsection{EM excitations in the momentum representation}
	
	In Section \ref{Sec:Doppler1}, the position-dependent electric field amplitudes $E_i(x_i,t_i)$ with $i={\rm A},{\rm B}$ were expressed as Fourier transforms of the momentum space field amplitudes $\widetilde E_i(k_i,t_i)$, characterised by the wave numbers $k_i$, which provided an alternative view of the classical Doppler transformation. The same transformation can also be performed for the blip annihilation operators $a_{s\mathsf{H}}(\chi_{\rm A})$. Doing so, we obtain annihilation operators $\tilde{a}_{s\lambda}(k_{\rm A}) $ with
	\begin{eqnarray}
		\label{F1}
		\tilde{a}_{s\lambda}(k_{\rm A}) &=& {1 \over \sqrt{2\pi}} \int_{-\infty}^{\infty} \text{d}\chi_{\rm A} \, {\rm e}^{-{\rm i}sk_{\rm A}\chi_{\rm A}} \, a_{s\lambda}(\chi_{\rm A})
	\end{eqnarray} 
	which describe field excitations characterised by a fixed wave number $k_{\rm A}$, direction of propagation $s$ and polarisation $\lambda$. Using commutation relation (\ref{Comm1}), we can show that the following commutation relation is satisfied:
	\begin{eqnarray}
		\label{Comm2}
		\left[\tilde{a}_{s\lambda}(k_{\rm A}), \tilde{a}^\dagger_{s'\lambda'}(k'_{\rm A})\right] &=& \delta_{ss'}\,\delta_{\lambda\lambda'}\,\delta(k_{\rm A}-k'_{\rm A}) \, .
	\end{eqnarray} 
	All other commutators are zero.  Due to the orthogonality of the $\tilde{a}_{s\lambda}(k_{\rm A})$ operators, $\tilde{a}_{s\lambda}(k_{\rm A})$ annihilates an excitation with a unique wave number $k_{\rm A}$.  The operator $\tilde{a}^\dagger_{s\lambda}(k_{\rm A})$ is the creation operator for this excitation. The inverse transformation of Eq.~(\ref{F1}) is given by
	\begin{eqnarray}
		\label{F2}
		a_{s\lambda}(\chi_{\rm A}) &=& {1 \over \sqrt{2\pi}} \int_{-\infty}^{\infty} \text{d}k_{\rm A} \, {\rm e}^{{\rm i}sk_{\rm A}\chi_{\rm A}} \, \tilde{a}_{s\lambda}(k_{\rm A})
	\end{eqnarray}
	which decomposes a single blip into a quantum superposition of monochromatic excitations for all $k_{\rm A} \in (-\infty, \infty)$.  As all possible values of $k_{\rm A}$ contribute to a localised excitation, the wavelength and the momentum of a single blip are completely undetermined.  Nevertheless, as the free EM field observables must propagate at the speed $c$ without any dispersion (like the corresponding solutions of Maxwell's equations), blip excitations must propagate in this way also.  As a result, the dynamics of single blips can be determined \cite{Hodgson:2021eyb}.	
	
	\section{A quantum picture of the relativistic Doppler effect}
	\label{Sec:4}
	
	Next we have a closer look at how Bob experiences the quantised EM field in his moving reference frame. Afterwards, we determine the relationship between Alice's and Bob's field observables using the classical field amplitude transformations derived in Section \ref{Sec:2B}. By taking into account that both Alice and Bob can express their field observables as a superposition of blips along both the $\chi_{\rm A} = x_{\rm A} - sct_{\rm A}$ and $\chi_{\rm B} = x_{\rm B} - sct_{\rm B}$ axes respectively, a local transformation is determined between blips in Bob's and Alice's reference frames.   
	
	\subsection{The Doppler effect in position space}
	
	In the following, we denote the annihilation and creation operators for a blip at a position $x_{\rm B}$ at an initial time $t_{\rm B} = 0$ in Bob's reference frame $b_{s\lambda}(x_{\rm B})$ and $b^\dagger_{s\lambda}(x_{\rm B})$ respectively with $s$ and $\lambda$ indicating again the direction of propagation and polarisation of the blip.  According to Bob, blips travel at the speed of light $c$ along the $x_{\rm B}$ axis and 
	\begin{eqnarray}
		\label{dynB}
		U_{\rm B}(t_{\rm B},0) \, b^\dagger_{s\lambda}(x_{\rm B}) \, U^\dagger_{\rm B}(t_{\rm B},0) &=& b^\dagger_{s\lambda}(x_{\rm B} -sct_{\rm B}) \, , ~~
	\end{eqnarray}
	in analogy to Eq.~(\ref{dyn}). Here $U_{\rm B}(t_{\rm B},0) $ denotes the time evolution operator of the EM field in Bob's reference frame. As in Section \ref{Sec:3}, constraining the blip annihilation operators in this way allows us to introduce annihilation operators $b_{s\lambda}(\chi_{\rm B})$ for blips at $x_{\rm B}$ and $t_{\rm B}$ with $\chi_{\rm B} = x_{\rm B} -sct_{\rm B}$.  Moreover 
	\begin{eqnarray}
		\label{Comm1B}
		\left[b_{s\lambda}(\chi_{\rm B}), b^\dagger_{s'\lambda'}(\chi'_{\rm B})\right] = \delta_{ss'}\,\delta_{\lambda\lambda'}\,\delta(\chi_{\rm B}-\chi'_{\rm B}) 
	\end{eqnarray} 
	in analogy to Eq.~(\ref{Comm1}). The blip operators in Bob's reference frame satisfy an identical set of commutation relations to Alice's operators, as one would expect from the principle of relativity.
	
	Entirely analogous to the generalisation of Alice's field observables, any non-local contributions of blips to Bob's field observables can be expressed in terms of their separation from Bob along the $\chi_{\rm B}$ axis.  Hence, we define Bob's field observables as
	\begin{eqnarray}
		\label{Efield2}
		E_{\rm B}(\chi_{\rm B}) &=& \sum_{s=\pm1} \int_{-\infty}^{\infty}\text{d}\chi'_{\rm B} \, c \, \mathcal{R}(\chi_{\rm B}-\chi'_{\rm B}) \, b_{s\mathsf{H}}(\chi'_{\rm B}) \notag \\
		&& + {\rm H.c.} \, , \notag \\
		B_{\rm B}(\chi_{\rm B}) &=& \sum_{s=\pm1} \int_{-\infty}^{\infty}\text{d}\chi'_{\rm B} \, s\,\mathcal{R}(\chi_{\rm B}-\chi'_{\rm B}) \, b_{s\mathsf{H}}(\chi'_{\rm B}) \notag \\
		&& + {\rm H.c.} 
	\end{eqnarray} 
	with $\chi_{\rm B} = x_{\rm B} -sct_{\rm B}$. Here $(x_{\rm B},t_{\rm B})$ are the spacetime coordinates of the point where the field measurement is made. The distribution $\mathcal{R}(\chi_{\rm B}-\chi'_{\rm B})$ in the equation above is the same as in Eq.~(\ref{Reg1}) but with $\chi_{\rm A}$ and $\chi'_{\rm A}$ replaced with $\chi_{\rm B}$ and $\chi_{\rm B}'$ respectively.  
	
	\begin{figure}
		\centering
		\includegraphics[width = 0.9\columnwidth]{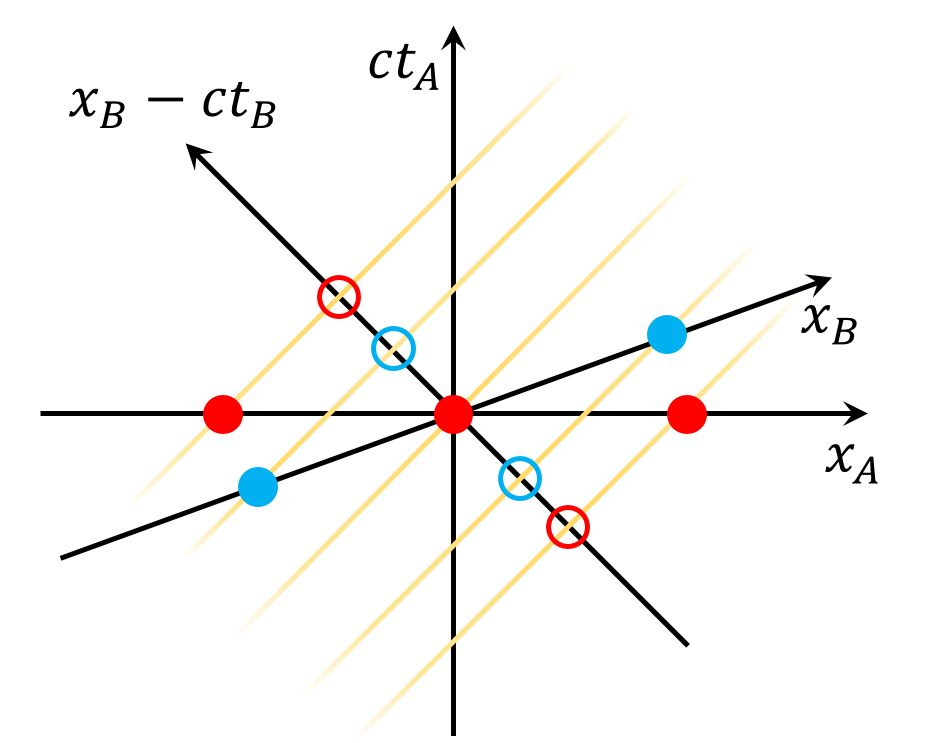}
		\caption{The diagram shows the contribution of right-propagating blips to Alice's and Bob's field observables at the origin. Blips distributed along the $x_{\rm A}$ axis (solid red) and along the $x_{\rm B}$ axis (solid blue) contribute equally to Alice's and Bob's field observables respectively. As blips at one point in spacetime can be identified with blips at all points along their world-lines (marked in yellow), blips distributed along the $\chi_{\rm B} = x_{\rm B}-ct_{\rm B}$ axis (hollows in red and blue) provide an equivalent contribution to Alice's and Bob's field observables as blips along the $x_{\rm A}$ axis and $x_{\rm B}$ axis on the same world-line.}
		\label{Fig:timeB}
	\end{figure}
	
	Although the same regularisation function is used by both Alice and Bob, as Alice measures the separation of blips along the $\chi_{\rm A}$ axis and Bob along the $\chi_{\rm B}$ axis, the contribution of blips to their field observables is different.  This is illustrated in Fig.~\ref{Fig:timeB}. In the diagram, Alice's blips (solid red) are distributed at regular intervals along the $x_{\rm A}$ axis.  Bob's blips (solid blue) are distributed at identical intervals along the $x_{\rm B}$ axis. Since $\mathcal{R}(x-x')$ is used for both observers, the two outermost red blips have the same contribution to Alice's field observables as the two outermost blue blips do to Bob's field observables.  By identifying the blips with their counterparts along the $x_{\rm B}-ct_{\rm B}$, however, we can see that the blips defined by Bob appear closer to Alice than her own.  As a result, the regularisation function used by Bob appears squeezed from Alice's point of view.  Conversely, for light propagating to the left the regularisation function appears stretched. As mentioned already above, events that occur simultaneously in Alice's reference frame are not simultaneous in Bob's reference frame and vice versa.    
	
	Now that the operators $a_{s\lambda}(\chi_{\rm A})$ and $b_{s\lambda}(\chi_{\rm B})$ have been defined by constructing field observables for the two observers (cf.~Eqs.~(\ref{Fields1}) and (\ref{Efield2})), a relationship between these operators can be found by imposing the field relation in Eq.~(\ref{2}). In order to compare the electric field observables $E_{\rm A}(\chi_{\rm A})$ and $E_{\rm B}(\chi_{\rm B})$, we now express $E_{\rm B}(\chi_{\rm B})$ in terms of Alice's coordinates $\chi_{\rm A}$. Combining Eqs.~(\ref{2}) and (\ref{Fields1}), we see that
	\begin{eqnarray}
		\label{Efield4}
		E_{\rm B}(\chi_{\rm B}) &=& \sum_{s=\pm1} \xi_{\rm BA} \int_{-\infty}^{\infty}\text{d}\chi'_{\rm A}\, c \, \mathcal{R}(\chi_{\rm A} - \chi'_{\rm A}) \, a_{s\mathsf{H}}(\chi'_{\rm A})  \notag \\
		&& + {\rm H.c.}
	\end{eqnarray} 
	with $\chi_{\rm B} = \gamma(1+s\beta) \, \chi_{\rm A}$. By substituting $\chi'_{\rm B} = \gamma(1+s\beta) \, \chi'_{\rm A}$ into Eq.(\ref{Efield4}) and using both Eq.~(\ref{fieldtransforms8}) and the explicit form of the regularisation function (\ref{Reg1}), we find that
	\begin{eqnarray}
		\label{Efield3}
		E_{\rm B}(\chi_{\rm B}) 
		&=& \sum_{s=\pm1} \left|\gamma(1-s\beta)\right|^{1/2} \int_{-\infty}^{\infty}\text{d}\chi'_{\rm B}\, c \, \mathcal{R}(\chi_{\rm B}-\chi_{\rm B}') \notag \\
		&& \times \, a_{s\mathsf{H}}(\gamma(1-s\beta) \chi_{\rm B}') + {\rm H.c.}
	\end{eqnarray} 
	By comparing this equation with Eq.~(\ref{Efield2}) we see that both expressions are only the same if 
	\begin{eqnarray}
		\label{blip1}
		b_{s\lambda}(\chi_{\rm B}) &=& \left[\gamma(1-s\beta)\right]^{1/2}\,a_{s\lambda}(\gamma(1-s\beta) \chi_{\rm B})
	\end{eqnarray}
	where the coordinates $\chi_{\rm A}$ and $\chi_{\rm B}$ define the same light-like trajectory in Alice's and Bob's coordinate systems respectively. One can easily check that the annihilation operators on both sides of the above transformation obey bosonic commutation relations. Hence the annihilation operators $a_{s\lambda}(\chi_{\rm A})$ and $b_{s\lambda}(\chi_{\rm B}) $ can be used interchangeably and a single blip in Alice's reference frame is observed by Bob as a single blip at exactly the same position in the spacetime diagram. 
	
	Moreover, the above result allows us to demonstrate that the total number of photons remains the same in both reference frames. In particular, using Eqs.~(\ref{trans1}) and (\ref{blip1}), one can check that
	\begin{eqnarray}
		\label{photon1}
		&& \hspace*{-1cm} \sum_{s=\pm1}\sum_{\lambda = {\sf H,V}}\int_{-\infty}^{\infty}\text{d}x_{\rm B}\;b_{s\lambda}^\dagger(x_{\rm B}) b_{s\lambda}(x_{\rm B})\nonumber\\
		&=& \sum_{s=\pm1}\sum_{\lambda = {\sf H,V}}\int_{-\infty}^{\infty}\text{d}x_{\rm A}\;a_{s\lambda}^\dagger(x_{\rm A}) a_{s\lambda}(x_{\rm A}) \, .
	\end{eqnarray} 
	In the local quantum picture of the Doppler shift which we present here, there is therefore no change to the particle nature of the EM field: a fixed number of local photons in Alice's reference frame also appears as an identical number of local photons in Bob's reference frame.  Hence, if both observers perform a linear optics experiment, for example, a Hong-Ou Mandel experiment \cite{HOM,HOM2} in which two identical photons approach a beam splitter from opposite sides, Alice and Bob both see both photons leaving the setup through the same output port. The dynamics of the quantised EM field is essentially the same in all inertial reference frames, as stated by Einstein's principle of relativity \cite{Einstein}.  The change in the amplitude of the blip operators by a factor of $\left[\gamma(1-s\beta)\right]^{1/2}$ in Eq.~(\ref{blip1}) is the direct result of Alice and Bob using different coordinates to describe the same spacetime point.  The relativistic Doppler effect is simply an immediate consequence of this fact.
	
	\subsection{The Doppler effect in the momentum representation}
	\label{Sec:IVB}
	
	For completeness, and since the relativistic Doppler effect is usually studied in momentum space \cite{Schachinger}, we finally have a closer look at the implications of the above equations on the momentum representation of the quantised EM field. In this representation, the electric and magnetic field observables are expressed in a basis of bosonic excitations with a definite frequency, polarisation and direction of propagation. Analogous to Alice's $\tilde{a}_{s\lambda}(k_{\rm A})$ operators defined in Eq.~(\ref{F1}), we now introduce a set of annihilation operators $\tilde{b}_{s\lambda}(k_{\rm B})$ with
	\begin{eqnarray}
		\label{F3}
		\tilde{b}_{s\lambda}(k_{\rm B}) &=& {1 \over \sqrt{2\pi}} \int_{-\infty}^{\infty} \text{d}\chi_{\rm B} \, {\rm e}^{-{\rm i}sk_{\rm B}\chi_{\rm B}}\,b_{s\lambda}(\chi_{\rm B}) 
	\end{eqnarray}
	in Bob's reference frame. Like the $\tilde{a}_{s\lambda}(k_{\rm A})$ operators, the $\tilde{b}_{s\lambda}(k_{\rm B})$ operators satisfy bosonic commutation relations:
	\begin{eqnarray}
		\label{Comm3}
		\left[\tilde{b}_{s\lambda}(k_{\rm B}), \tilde{b}^\dagger_{s'\lambda'}(k'_{\rm B})\right] = \delta_{ss'}\,\delta_{\lambda\lambda'}\,\delta(k_{\rm B}-k'_{\rm B}) \, .
	\end{eqnarray}
	Here $k_{\rm B}c$ is the frequency measured by Bob with respect to the time $t_{\rm B}$.
	
	In the position representation, the blip operators $a_{s\lambda}(\chi_{\rm A})$ and $b_{s\lambda}(\chi_{\rm B})$ defined in Alice's and Bob's reference frames respectively satisfy the transformation in Eq.~(\ref{blip1}). In the following, we derive an analogous relationship between the corresponding momentum space annihilation operators. Substituting Eq.~(\ref{blip1}) into the right hand side of Eq.~(\ref{F3}) and again taking into account that $\chi_{\rm B} = \gamma(1+s\beta)\, \chi_{\rm A}$ (cf.~Eq.~(\ref{trans1})), we find that
	\begin{eqnarray}
		\label{momtrans1}
		\tilde b_{s\lambda}(k_{\rm B}) &=& {1 \over \sqrt{2\pi}} \left[\gamma(1+s\beta)\right]^{1/2} \int_{-\infty}^{\infty} \text{d}\chi_{\rm A} \notag \\
		&& \times {\rm e}^{-{\rm i}s \gamma(1+s\beta) k_{\rm B} \chi_{\rm A}} \,a_{s\lambda}(\chi_{\rm A}) \, .
	\end{eqnarray}
	Then, using Eq.~(\ref{F2}), we obtain
	\begin{eqnarray}
		\label{momtrans2}
		\tilde b_{s\lambda}(k_{\rm B}) 
		&=& {1 \over 2\pi} \left[\gamma(1+s\beta)\right]^{1/2}\int_{-\infty}^{\infty} \text{d}\chi_{\rm A}\int_{-\infty}^{\infty} \text{d}k_{\rm A} \notag \\
		&& \times {\rm e}^{{\rm i}s [k_{\rm A}-\gamma(1+s\beta) k_{\rm B}] \chi_{\rm A}} \, \tilde a_{s\lambda}(k_{\rm A}) \,.
	\end{eqnarray} 
	After performing the $\chi_{\rm A}$ integration, which yields a Dirac delta function in $k_{\rm A}-\gamma(1+s\beta) k_{\rm B}$, and then the $k_{\rm A}$ integration, we see that
	\begin{eqnarray}
		\label{momtrans2final}
		\tilde b_{s\lambda}(k_{\rm B}) &=& \left[\gamma(1+s\beta)\right]^{1/2} \, \tilde a_{s\lambda}(\gamma(1+s\beta)k_{\rm B}) \, .
	\end{eqnarray} 
	As for Eq.~(\ref{blip1}), the annihilation operators on both sides of this equation obey bosonic commutation relations. The above result therefore demonstrates that the $\tilde b_{s\lambda}(k_{\rm B}) $ and $\tilde a_{s\lambda}(k_{\rm A}) $ can be used interchangeably so long as Alice's and Bob's wave numbers are such that
	\begin{eqnarray}
		\label{oma}
		k_{\rm A} &=& \gamma(1+s\beta) \, k_{\rm B} \, . 
	\end{eqnarray} 
A single monochromatic photon with wave number $k_{\rm B}$ observed by Bob therefore appears as a single monochromatic photon to Alice, but its wave number is altered as described by the above relation. The corresponding frequency transformation is in complete agreement with the classical Doppler shift derived in Section \ref{Sec:Doppler1} (cf.~Eq.~(\ref{frequency4})).  
	
	In addition to a frequency transformation, there is also a change in the amplitude of monochromatic excitations. This amplitude change occurs as the transformation between excitations with a shift between sharp frequencies cannot be unitary, as shown in Ref.~\cite{Bruschi1} for gravitationally redshifted photons. It was also shown in this reference, however, that a unitary transformation can be constructed for realistic photon operators when a suitable transformation for the frequency distribution of the photon is introduced \cite{Bruschi1, Bruschi5}. Whilst the transformation given in Eq.~(\ref{momtrans2final}) is very simple, it shows clearly that excitations do not have the same momenta in all reference frames. In particular, if Alice detects an excitation in the $k_{\rm A}$ mode only, this mode will be empty according to Bob.  Bob, however, will detect an equal number of excitations in his $k_{\rm B} = \gamma(1-s\beta)k_{\rm A}$ mode.  This result is very different to that in the position representation where a local photon is viewed as a local photon at the same position in the spacetime diagram by all inertial observers. The only difference is that different observers use different coordinates to characterise this point.
	
	\section{Conclusions}
	\label{Sec:5}
	
	This paper offers an alternative perspective on the relativistic Doppler effect, which is usually referred to only in momentum space and discussed in terms of frequency, wavelength and amplitude changes of wave packets of light when measured in different inertial reference frames. In this paper, we study the relativistic Doppler effect in position space using the spacetime coordinates $\chi_{\rm A}=x_{\rm A}-sct_{\rm A}$ and $\chi_{\rm B}=x_{\rm B}-sct_{\rm B}$ of two inertial observers: Alice in the stationary frame and Bob in a frame moving with constant velocity $v_{\rm B}$ with respect to Alice. This alternative approach allows us to accommodate spatial and time translational symmetries in a relatively straightforward way. In addition, we take advantage of energy conservation and the principle of relativity.
	
	For example, symmetry arguments and the principle of relativity are used to show that local electric field amplitudes seen by Alice and Bob only differ by a constant factor which we denote $\xi_{\rm AB}$ and $\xi_{\rm BA}$ respectively. Energy conservation can be used to calculate these factors as a function of the propagation direction $s$ and the velocity of Bob's reference frame $v_{\rm B}$ with respect to Alice's frame. For simplicity, we assume here that both observers are stationary in their respective coordinate systems and place them at the origin. When transforming our local description of the relativistic Doppler effect into momentum space, we recover the usual predictions which shows that our approach is consistent with the findings of other Authors. 
	
	Sections \ref{Sec:3} and \ref{Sec:4} concentrate on the local description of the quantized EM field for light propagation in the $1+1$ dimensional Minkowski spacetime to obtain a quantum picture of the relativistic Doppler effect. Our aim here is to identify the relationship between the quantum states of a wave packet of light seen by Alice and Bob. Our main result is the straightforward relationship between the annihilation operators $a_{s \lambda} (\chi_{\rm A})$ and  $b_{s \lambda} (\chi_{\rm B})$ used by Alice and Bob for the description of local excitations\textemdash so-called blips\textemdash of the quantised EM field. When considering the same point in the spacetime diagram, i.e.~when $\chi_{\rm A}$ and $\chi_{\rm B}$ depend on each other as stated in Eq.~(\ref{trans1}), both observers measure the same number of field excitations. 
	
	Although the classical electric and magnetic fields undergo a change in amplitude (cf.~Eq.~(\ref{2})), the photon intensity remains the same, which is an important result of quantum physics.  In the photoelectric effect, for example, it is the photon number and not the field intensity that determines the number of emitted electrons. Moreover, we conclude that the relativistic Doppler effect is not a quantum effect but simply the result of Alice and Bob using different spacetime coordinates and experiencing space and time differently, while the speed of light is the same in all inertial reference frames \cite{Otting1939}. Furthermore, as was shown in Section \ref{Sec:IVB}, although the total number of excitations is conserved, in the momentum representation Alice and Bob observe different numbers of excitations in each $k$ mode.
	
	The results of this paper might have applications in different areas of physics, including quantum communication \cite{Bruschi2, Perdigues2008} and relativistic quantum information \cite{Bruschi3, Ralph:2011hp,Friis:2012cx,Ursin2009, Alsing:2012wf,Rideout:2012jb}. Our results might also have some implications for ongoing discussions into the basic assumptions of relativity theory, since it 
	implies that the local description of the quantised EM field is equivalent in both Alice's and Bob's reference frames. A different set of assumptions would lead to an alternative relationship between blips in these two reference frames.  Our results on the transverse Doppler effect may aid experimental verifications of the reality of length contraction and time dilation. In the current study, the concentration was on a well-expected result caused by the transformation between the stationary frame and a moving frame in a classical and relativistic representation in order to explore it for blips as well. In the future, our approach can be used to study more complicated situations and systems like an accelerating reference frame. \\[0.5cm]
	{\em Acknowledgement.} D. H. acknowledges financial support from the UK Engineering and Physical Sciences Research Council EPSRC [grant number EP/W524372/1]. In addition, the authors thank Basil M. Altaie for helpful comments and stimulating discussions.
	
	\bibliographystyle{apsrev}
	
\end{document}